\documentclass[amsmath,amssymb,nofootinbib,prd,preprint,superscriptaddress]{revtex4-1}
\pdfoutput=1

\usepackage[allcolors=blue,colorlinks=true]{hyperref}
\usepackage{xspace}
\usepackage{tikz}
\usetikzlibrary{arrows.meta}

\newcommand{\br}[1]{\left[#1\right]}
\newcommand{\pa}[1]{\left(#1\right)}
\newcommand{\ads}{$AdS_2$\xspace}
\newcommand{\sltwo}{$\mathsf{SL}(2)$\xspace}
\newcommand{\p}{\partial}

\DeclareMathOperator{\sgn}{sgn}

\begin{document}

\title{Extreme Black Hole Anabasis}

\author{Shahar Hadar}
\email{shaharhadar@g.harvard.edu}
\affiliation{Center for the Fundamental Laws of Nature, Harvard University, Cambridge, MA 02138, USA}
\author{Alexandru Lupsasca}
\email{lupsasca@princeton.edu}
\affiliation{Princeton Gravity Initiative, Princeton University, Princeton, NJ 08544, USA}
\author{Achilleas P. Porfyriadis}
\email{porfyr@g.harvard.edu}
\affiliation{Center for the Fundamental Laws of Nature, Harvard University, Cambridge, MA 02138, USA}
\affiliation{Black Hole Initiative, Harvard University, Cambridge, MA 02138, USA}

\begin{abstract}
We study the \sltwo transformation properties of spherically symmetric perturbations of the Bertotti-Robinson universe and identify an invariant $\mu$ that characterizes the backreaction of these linear solutions.  The only backreaction allowed by Birkhoff's theorem is one that destroys the $AdS_2\times S^2$ boundary and builds the exterior of an asymptotically flat Reissner-Nordstr\"om black hole with $Q=M\sqrt{1-\mu/4}$.  We call such backreaction with boundary condition change an \emph{anabasis}.  We show that the addition of linear anabasis perturbations to Bertotti-Robinson may be thought of as a boundary condition that defines a \emph{connected} $AdS_2\times S^2$.  The connected \ads is a nearly-\ads with its \sltwo broken appropriately for it to maintain connection to the asymptotically flat region of Reissner-Nordstr\"om.  We perform a backreaction calculation with matter in the connected $AdS_2\times S^2$ and show that it correctly captures the dynamics of the asymptotically flat black hole.
\end{abstract}

\maketitle

\section{Introduction}

Birkhoff's theorem in four dimensions tells us that all spherically symmetric spacetimes with vanishing Ricci tensor are static and therefore described by the Schwarzschild metric.  The theorem extends to the Einstein-Maxwell equations with the Schwarzschild solution replaced by the Reissner-Nordstr\"om one.  On the other hand, in the Einstein-Maxwell theory another spherically symmetric solution of importance is the Bertotti-Robinson universe with metric given by the direct product of \ads with a two-sphere.  This is consistent with Birkhoff's theorem because Bertotti-Robinson agrees with the near-horizon of extreme and near-extreme Reissner-Nordstr\"om, and the theorem is only a local statement.\footnote{The precise statement of Birkhoff's theorem is that a $C^2$ solution of the Einstein (resp. Einstein-Maxwell) equations which is spherically symmetric in an open set $\mathcal{V}$ is locally equivalent to part of the maximally extended Schwarzschild (resp. Reissner-Nordstr\"om) solution in $\mathcal{V}$ (see, e.g., \cite{Hawking:1973uf}).}  Reissner-Nordstr\"om is asymptotically flat while Bertotti-Robinson is asymptotically $AdS_2\times S^2$. 

If one considers the linearized Einstein-Maxwell equations around Reissner-Nordstr\"om, then within spherical symmetry one finds a two-parameter family of solutions parametrized by the change in mass $\delta M$ and charge $\delta Q$ relative to the background.  In other words, such linearized perturbations are moving towards other Reissner-Nordstr\"om solutions.  On the other hand, if one considers the linearized Einstein-Maxwell equations around Bertotti-Robinson then one finds a four-parameter family of solutions.  Why is that?  One may identify two of the four parameters with $\delta M\pm\delta Q$ and derive the corresponding solutions from appropriate near-horizon limits of the linearized solutions around Reissner-Nordstr\"om.  When $\delta M=\delta Q$ this results in a linearized solution around Bertotti-Robinson that respects the \sltwo symmetry associated with the \ads factor of the background.  This \sltwo-preserving linear solution is asymptotically $AdS_2\times S^2$ and moves towards another Bertotti-Robinson.  When $\delta M\neq\delta Q$ the near-horizon limit produces a linear solution around Bertotti-Robinson that breaks its \sltwo symmetry.  Therefore acting with the background's \sltwo isometries we may obtain two additional linear solutions.  The \sltwo-breaking triplet of solutions are not asymptotically $AdS_2\times S^2$ and it is well known that the backreaction of these solutions destroys the \ads boundary \cite{Maldacena:1998uz}.  In the past, this has led to the slogans that ``\ads has no dynamics'' or that ``\ads admits no finite energy excitations.''  While not incorrect, these slogans are true only as long as one insists on asymptotically \ads boundary conditions.

In this paper, we study the \sltwo-breaking triplet solutions with boundary conditions that allow the near-horizon $AdS_2\times S^2$ throat to maintain its connection with the exterior asymptotically flat Reissner-Nordstr\"om.  In this context, we first identify an \sltwo-invariant quantity $\mu$ associated with any solution in the triplet.  Then we show that when $\mu=0$ the corresponding solution may be thought of as beginning to build the asymptotically flat region of extreme Reissner-Nordstr\"om starting from its $AdS_2\times S^2$ throat.  When $\mu>0$ the corresponding solution is beginning to build the asymptotically flat region of near-extreme Reissner-Nordstr\"om with $Q=M\sqrt{1-\mu/4}$.  In other words, when $\mu\geq0$ the backreaction of these \sltwo-breaking linear solutions makes sense provided we allow the boundary condition change that leads to an asymptotically flat nonlinear solution.  We call such backreaction with boundary condition change an \emph{anabasis}---an adventure of climbing out of the black hole throat into the weak gravity regime.

Next, consider perturbing an extreme Reissner-Nordstr\"om black hole using a spherically symmetric infalling matter source with energy-momentum tensor of order $\epsilon\ll1$.  Generically, one expects the fully backreacted nonlinear endpoint of this perturbation to be a near-extreme Reissner-Nordstr\"om with $Q=M\sqrt{1-\mathcal{O}(\epsilon)}$ \cite{Murata:2013daa}.  To leading order in $\epsilon$, the initial and final states only differ in their near-horizon region.  More precisely, the near-horizon throat geometry remains locally $AdS_2\times S^2$ before as well as after the perturbation but the associated \sltwo symmetry breaking induced by the gluing to the exterior region is different in the extreme and near-extreme cases.  We define the \emph{connected} $AdS_2\times S^2$ throat as the geometry obtained by the addition of anabasis perturbations.  This may also be thought of as a boundary condition for the backreaction calculation.  We show that this leads to a consistent backreaction calculation in $AdS_2\times S^2$ that captures the dynamics of the asymptotically flat black hole.

In Section~\ref{sec: BR pertns}, we derive the spherically symmetric perturbations of the Bertotti-Robinson universe and study their \sltwo transformation properties.  In particular, we identify the invariant quantity $\mu$ associated with each \sltwo-breaking perturbation.  Section~\ref{sec: anabasis pertns} singles out standard Poincar\'e and Rindler anabasis perturbations responsible for building the exterior in extreme and near-extreme Reissner-Nordstr\"om, respectively.  Section~\ref{sec: general PtoR  transformation} presents the general transformation from Poincar\'e to Rindler $AdS_2\times S^2$.  In Section~\ref{sec: backreaction calcn}, we do a backreaction calculation for a pulse of energy $\epsilon$ in the connected $AdS_2\times S^2$ throat of a Reissner-Nordstr\"om black hole. Section~\ref{sec: conclusion} contains further discussion of our work, especially in relation to the AdS/CFT correspondence and models of two-dimensional dilaton gravity in $AdS_2$.

\section{Perturbations of Bertotti-Robinson}
\label{sec: BR pertns}

The Einstein-Maxwell equations in four dimensions read ($G=c=1$)
\begin{align}
	\label{Einstein-Maxwell}
	R_{\mu\nu}=8\pi T_{\mu\nu}\,,\quad
	\nabla^\mu F_{\mu\nu}=0\,,
\end{align}
with $4\pi T_{\mu\nu}=F_{\mu\rho}F_\nu^{~\rho}-\frac{1}{4}g_{\mu\nu}F_{\rho\sigma}F^{\rho\sigma}$, and $F_{\mu\nu}=\p_\mu A_\nu-\p_\nu A_\mu$.  The Bertotti-Robinson universe
\begin{align}
	\label{AdS2xS2}
	\frac{1}{M^2}ds^2=-r^2dt^2+\frac{dr^2}{r^2}+d\Omega^2\,,\quad
	A_t=Mr\,,
\end{align}
is a spherically symmetric conformally flat exact solution with uniform electromagnetic field $F_{rt}=M$.  Clearly the Bertotti-Robinson metric is a direct product $AdS_2\times S^2$.  From now on, we set $M=1$ and restore it only when beneficial for clarity.

Consider the most general spherically symmetric perturbation
\begin{align}
\label{pertn ansatz}
\begin{aligned}
	h_{\mu\nu}=&
	\begin{pmatrix}
		h_{tt}(t,r) & h_{tr}(t,r) & 0 & 0 \\
		& h_{rr}(t,r) & 0 & 0 \\
		& & h_{\theta\theta}(t,r) & 0 \\
		& &	& h_{\theta\theta}\sin^2{\theta}
	\end{pmatrix}
	\,,\\
	a_\mu=&
	\begin{pmatrix}
		a_t(t,r) & a_r(t,r) & 0 & 0 
	\end{pmatrix}
	\,.
\end{aligned}
\end{align}
The linearized Einstein-Maxwell equations are invariant under the gauge transformations
\begin{align}
	\label{diffeo-gauge transfn}
	h_{\mu\nu}\to h_{\mu\nu}+\mathcal{L}_\xi g_{\mu\nu}\,,\quad
	a_\mu\to a_\mu+\mathcal{L}_\xi A_\mu+\nabla_\mu\Lambda\,,
\end{align}
for any vector field $\xi$ and scalar function $\Lambda$.  Within the spherically symmetric ansatz, we may use this gauge freedom, with appropriate $\xi=\xi^t(t,r)\p_t+\xi^r(t,r)\p_r\,,\Lambda=\Lambda(t,r)$, to set
\begin{align}
	\label{gauge choice}
	h_{tt}=h_{rr}=a_t=0\,,
\end{align}
and remove from $h_{tr}$ any addition of the form $h_{tr}=c_1(r)+c_2(t)/r$ for arbitrary $c_1$, $c_2$.  Note, however, that for perturbations around the Bertotti-Robinson solution $h_{\theta\theta}$ is gauge invariant.  Therefore, all physical information for perturbations of Bertotti-Robinson is contained in $h_{\theta\theta}$.

We find that the most general solution to the linearized Einstein-Maxwell equations around Bertotti-Robinson is given by
\begin{align}
	\label{hthth soln}
	h_{\theta\theta}&=\Phi_0+ar+brt+cr\pa{t^2-1/r^2}\,,\\
	\label{htr soln}
	h_{tr}&=-\frac{1}{2}rt\br{\Phi_0+2ar+brt+\frac{2}{3}cr\pa{t^2-9/r^2}}\,,\\
	\label{ftr soln}
	f_{tr}&=\p_ta_r
	=h_{\theta\theta}-{\Phi_0/2}\,.
\end{align}
Thus, the spherically symmetric perturbations of Bertotti-Robinson are a four-parameter family of solutions parametrized by the constants $\Phi_0$, $a$, $b$, $c$.

\subsection{\texorpdfstring{\boldmath\sltwo}{SL(2)} transformations and invariants}
\label{subsection: SL(2)}

The background \eqref{AdS2xS2} is invariant under the \sltwo transformations associated with the \ads factor:
\begin{align}
	&H(\alpha):\qquad
	t\to t+\alpha\,,\\
	&D(\beta):\qquad
	t\to t/\beta\,,\quad
	r\to\beta r\,,\\
	&K(\gamma):\qquad
	t\to\frac{t-\gamma\pa{t^2-1/r^2}}{1-2\gamma t+\gamma^2\pa{t^2-1/r^2}}\,,\quad
	r\to r\br{1-2\gamma t+\gamma^2\pa{t^2-1/r^2}}\,.
\end{align}
Here $H(\alpha)$, $D(\beta)$, $K(\gamma)$ are the time translations, dilations, and special conformal transformations for real parameters $\alpha$, $\beta$, $\gamma$ with $\beta>0$.  The special conformal coordinate transformation must be followed by a gauge field transformation $A\to A+d\ln\frac{r(t-1/\gamma)+1}{r(t-1/\gamma)-1}$.

The \sltwo invariance of the background implies that if we act with an \sltwo transformation on any of the solutions (\ref{hthth soln}--\ref{ftr soln}) we will obtain another solution to the linearized Einstein-Maxwell equations around the same background.  Note, however, that the \sltwo transformations do not necessarily preserve the gauge \eqref{gauge choice}.  Fortunately, as we have previously emphasized, $h_{\theta\theta}$ is gauge invariant and therefore uniquely labels each physically distinct solution in every gauge.

The four-parameter solution \eqref{hthth soln} consists of an \sltwo-invariant solution $\Phi_0$ together with the \sltwo-breaking triplet
\begin{align}
	\label{Phi soln}
	\Phi=ar+brt+cr\pa{t^2-1/r^2}\,.
\end{align}
Clearly, the \sltwo-invariant solution $\Phi_0$ corresponds to a rescaling of \eqref{AdS2xS2} by $M\to M+\delta M$ with $\Phi_0=2M\,\delta M$.  In the remainder of the paper, we will focus on the \sltwo-breaking solutions $\Phi$.

The action of the \sltwo transformations on $\Phi$ is given by
\begin{align}
	&H(\alpha):\qquad
	a\to a+b\alpha+c\alpha^2\,,\quad
	b\to b+2c\alpha\,,\quad
	c\to c\,,\label{H transfn}\\
	&D(\beta):\qquad
	a\to a\beta\,,\quad
	b\to b\,,\quad
	c\to c/\beta\,,\label{D transfn}\\
	&K(\gamma):\qquad
	a\to a\,,\quad
	b\to b-2a\gamma\,,\quad
	c\to c-b\gamma+a\gamma^2\,.\label{K transfn}
\end{align}
Using the above, we identify the following \sltwo invariant
\begin{align}
	\label{mu defn}
	\mu=b^2-4ac\,.
\end{align}
Moreover, we note that for $\mu<0$ we have $\sgn{a}=\sgn{c}\neq0$ being an additional \sltwo invariant, while for $\mu=0$ it is $\sgn(a+c)$ that is also \sltwo-invariant.

From (\ref{H transfn}--\ref{K transfn}), we see that each of the \sltwo transformations preserves one and only one of the three solutions in \eqref{Phi soln}.  If we think of $(a,b,c)$ as charges associated with the corresponding \sltwo transformations, then $\mu$ may be identified with the quadratic \sltwo Casimir, which is invariant under the group action.  In fact, under a general \sltwo isometry, $(a,b,c)$ transforms in the coadjoint representation, thereby generating a coadjoint orbit.  The invariant $\mu$ labels the coadjoint orbits of \sltwo, which come in three types (elliptic, parabolic, and hyperbolic) according to the sign of $\mu$.

Before we end this section, let us fix two standard choices for the general solution \eqref{Phi soln}.  For $\mu>0$ one may always find an \sltwo transformation that will set
\begin{align}
	\label{Phi mu>0 standard}
	\Phi=-\sqrt{\mu}\,rt\,,\quad
	\mu>0\,.
\end{align}
Similarly, for $\mu=0$ and $\sgn (a+c)=1$ one may set
\begin{align}
	\label{Phi mu=0 standard}
	\Phi=2r\,,\quad
	\mu=0\,,\,
	\sgn (a+c)=1\,.
\end{align}
Specifically, when $\mu>0$ we may get to $\Phi=-\sqrt{\mu}\,rt$ by acting on \eqref{Phi soln} with the following series of \sltwo transformations:
\begin{align}
	&H\pa{\frac{a}{\sqrt{b^2-4ac}}}\circ K\pa{\frac{b+\sqrt{b^2-4ac}}{2a}}\,,
	&&\text{for}\quad
	a\neq 0\,,\\
	&K(\gamma)\circ H\pa{1/\gamma}\circ K(\gamma)\circ K\pa{\frac{c}{b}}\,,
	&&\text{for}\quad
	a=0\,,b>0\,,\\
	&K\pa{\frac{c}{b}}\,,
	&&\text{for}\quad a=0\,,b<0\,.
\end{align}
Likewise, when $\mu=0$ we may get to $\Phi=\sgn(a+c)\,2r$ by acting on \eqref{Phi soln} with the following series of \sltwo transformations:
\begin{align}
	&D\pa{\frac{2}{|a|}}\circ K\pa{\frac{b}{2a}}\,,
	&&\text{for}\quad
	abc\neq0\,,\\
	&D\pa{\frac{2}{\gamma^2|c|}}\circ K\pa{1/\gamma}\circ H(\gamma)\,,
	&&\text{for}\quad
	b=a=0\,,\\
	&D\pa{\frac{2}{|a|}}\,,
	&&\text{for}\quad
	b=c=0\,.
\end{align}

\section{Anabasis and the connected throat}
\label{sec: anabasis pertns}

The Bertotti-Robinson solution \eqref{AdS2xS2} may be derived from near-horizon near-extremality scalings of the Reissner-Nordstr\"om black hole solution of mass $M$ and charge $Q$, with outer/inner horizons at $r_\pm=M\pm\sqrt{M^2-Q^2}$,
\begin{align}
	\label{RN hatted metric}
	ds^2=-\pa{1-\frac{2M}{\hat{r}}+\frac{Q^2}{\hat{r}^2}}d\hat{t}^2+\pa{1-\frac{2M}{\hat{r}}+\frac{Q^2}{\hat{r}^2}}^{-1}d\hat{r}^2+\hat{r}^2d\Omega^2\,,\quad
	\hat{A}_{\hat{t}}=-\frac{Q}{\hat{r}}\,.
\end{align}
This implies that the leading corrections in these scaling limits are, by construction, solutions of the linearized Einstein-Maxwell equations around Bertotti-Robinson.  There are two essentially distinct scaling limits of the black hole exterior that yield the Bertotti-Robinson solution.

The first scaling limit is most simply described by setting $Q=M$,
making the coordinate and gauge transformation,
\begin{align}
	r=\frac{\hat{r}-M}{\lambda M}\,,\quad
	t=\frac{\lambda\hat{t}}{M}\,,\quad
	A=\hat{A}+d\hat{t}\,,
\end{align}
to obtain
\begin{align}
	\label{ERN lambda metric}
	\frac{1}{M^2}ds^2=-\pa{\frac{r}{1+\lambda r}}^2dt^2+\pa{\frac{r}{1+\lambda r}}^{-2}dr^2+\pa{1+\lambda r}^2d\Omega^2\,,\quad
	A_t=M\frac{r}{1+\lambda r}\,,
\end{align}
and then taking the limit $\lambda\to 0$.  At order $\mathcal{O}(1)$ this produces exactly \eqref{AdS2xS2}.  The leading correction is of order $\mathcal{O}(\lambda)$ and it is given by
\begin{align}
	\label{2r soln from limit}
	h_{tt}=2r^3\,,\quad
	h_{rr}=2/r\,,\quad
	h_{\theta\theta}=2r\,,\quad
	f_{rt}=-2r\,.
\end{align}
By construction, this solves the linearized Einstein-Maxwell equations around \eqref{AdS2xS2}. 

Comparing the gauge invariant $h_{\theta\theta}$ in the above with \eqref{hthth soln} we see that this is the $\Phi=2r$ solution.\footnote{Indeed, one may align \eqref{2r soln from limit} with the $a=2$, $b=c=\Phi_0=0$ solution (\ref{hthth soln}--\ref{ftr soln}) by adjusting the gauge via \eqref{diffeo-gauge transfn} with $\xi=2rt\p_t-r^2\p_r\,,\Lambda=0$.}  Hence the \sltwo-breaking $\mu=0$ solution $\Phi=2r$ may be thought of as beginning to build the asymptotically flat region of an extreme Reissner-Nordstr\"om starting from its near-horizon Bertotti-Robinson throat.  In other words, the nonlinear solution obtained from the $\mu=0$ perturbation of $AdS_2\times S^2$, when backreaction is fully taken into account in the Einstein-Maxwell theory, is the extreme Reissner-Nordstr\"om black hole.

The second scaling limit is described by setting $Q=M\sqrt{1-\lambda^2\kappa^2}$, making the coordinate and gauge transformation,
\begin{align}
	\rho=\frac{\hat{r}-r_+}{\lambda r_+}\,,\quad
	\tau=\frac{\lambda \hat{t}}{M}\,,\quad
	A=\hat{A}+d\hat{t}\,,
\end{align}
to obtain
\begin{align}
\label{NERN lambda metric}
\begin{aligned}
	\frac{1}{M^2}ds^2&=-\frac{\rho(\rho+2\kappa+\lambda\kappa\rho)}{(1+\lambda\kappa)(1+\lambda\rho)^2}d\tau^2+\frac{(1+\lambda\kappa)^3(1+\lambda\rho)^2}{\rho(\rho+2\kappa+\lambda\kappa\rho)}d\rho^2+(1+\lambda\kappa)^2(1+\lambda\rho)^2d\Omega^2\,,\\
	A_\tau&=\frac{M}{\lambda}\pa{1-\sqrt{\frac{1-\lambda\kappa}{1+\lambda\kappa}}\frac{1}{1+\lambda\rho}}\,,
\end{aligned}
\end{align}
and then taking the limit $\lambda\to0$.  At order $\mathcal{O}(1)$ this produces
\begin{align}
	\label{NAdS2xS2}
	\frac{1}{M^2}ds^2=-\rho(\rho+2\kappa)d\tau^2+\frac{d\rho^2}{\rho(\rho+2\kappa)}+d\Omega^2\,,\quad
	A_\tau=M(\rho+\kappa)\,.
\end{align}
The leading correction is of order $\mathcal{O}(\lambda)$ and it is given by
\begin{align}
	\label{2(r+k) soln from limit}
	h_{\tau\tau}=2\rho(\rho+\kappa)^2\,,\quad 
	h_{\rho\rho}=\frac{2\pa{\rho^2+3\kappa^2+3\kappa\rho}}{\rho(\rho+2\kappa)^2}\,,\quad 
	h_{\theta\theta}=2(\rho+\kappa)\,,\quad 
	f_{\rho\tau}=-2\rho-\kappa\,.
\end{align}
By construction, this solves the linearized Einstein-Maxwell equations around \eqref{NAdS2xS2}.

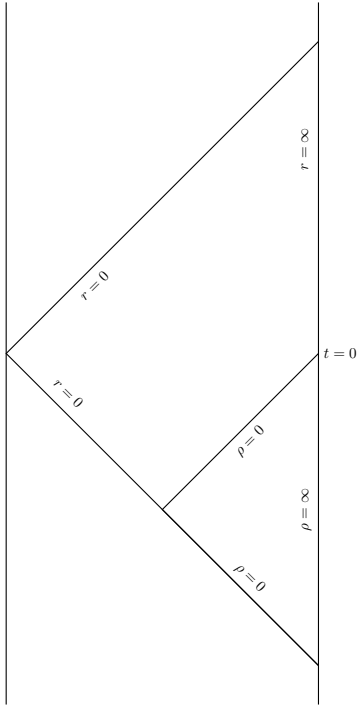
\begin{figure}[!ht]
	\centering
	\resizebox{!}{0.4\textheight}
	{
	\begin{tikzpicture}
		\draw[thick] (0,-1) -- (0, 17);
		\draw[thick] (-8,-1) -- (-8, 17);

		\draw[thick] (0,0) to (-8, 8) to (0,16);

		\draw[thick] (0,0) to (-4, 4) to (0,8);

		\draw (0, 8) node[anchor=west]{$t=0$};

		\draw (-5.75, 9.75) node[rotate=45]{$r=0$};
		\draw (-6.4, 7) node[rotate=-45]{$r=0$};
		\draw (-0.35, 13.25) node[rotate=90]{$r=\infty$};

		\draw (-1.75, 5.75) node[rotate=45]{$\rho=0$};
		\draw (-1.75,2.25) node[rotate=-45]{$\rho=0$};
		\draw (-0.3, 4) node[rotate=90]{$\rho=\infty$};
	\end{tikzpicture}
	}
	\caption{Penrose diagram of $AdS_2\times S^2$ with relative placement of the Poincar\'e patch \eqref{AdS2xS2} and the Rindler patch \eqref{NAdS2xS2} according to the transformation \eqref{slow nearNHEK transfn}.}
	\label{fig1}
\end{figure}

Locally, the $\mathcal{O}(1)$ results of the two scaling limits we have considered [Eqs.~\eqref{AdS2xS2} and \eqref{NAdS2xS2}] are diffeomorphic---they are both the Bertotti-Robinson universe.  Indeed, the coordinate transformation \cite{Spradlin:1999bn}
\begin{align}
	\label{slow nearNHEK transfn}
	\tau=-\frac{1}{2\kappa}\ln\pa{t^2-1/r^2}\,,\quad 
	\rho=-\kappa(1+rt)\,,
\end{align}
together with $A\to A-d\Lambda\,,\Lambda=\frac{1}{2}\ln\frac{rt-1}{rt+1}=-\frac{1}{2}\ln\frac{\rho}{\rho+2\kappa}$ maps \eqref{NAdS2xS2} to \eqref{AdS2xS2}.  Globally, on the Penrose diagram of $AdS_2\times S^2$, the coordinates in \eqref{AdS2xS2} cover a Poincar\'e patch while the coordinates in \eqref{NAdS2xS2} cover a Rindler patch.  The transformation \eqref{slow nearNHEK transfn} situates the two patches relative to each other as shown in Fig.~\ref{fig1}.  It follows that under this transformation the leading $\mathcal{O}(\lambda)$ correction \eqref{2(r+k) soln from limit} transforms to a solution of the linearized Einstein-Maxwell equations around \eqref{AdS2xS2}.  Comparing the gauge invariant $h_{\theta\theta}=2(\rho+\kappa)=-2\kappa rt$ with \eqref{hthth soln}, we see that this is the $\Phi=-2\kappa rt$ solution.  Hence the \sltwo-breaking $\sqrt{\mu}=2\kappa$ solution $\Phi=-2\kappa rt$ may be thought of as beginning to build the asymptotically flat region of a near-extreme Reissner-Nordstr\"om starting from its near-horizon Bertotti-Robinson throat.  In other words, the nonlinear solution obtained from the $\mu>0$ perturbation of $AdS_2\times S^2$, when backreaction is fully taken into account in the Einstein-Maxwell theory, is the near-extreme Reissner-Nordstr\"om black hole with $Q=M\sqrt{1-\mu/4}$.

We end this section by bringing attention to the fact that the anabasis solutions $\Phi=2r$ and $\Phi=2(\rho+\kappa)$ that begin to build the asymptotically flat black hole exteriors are positive
\begin{align}
	\label{Phi>0}
	\Phi>0\,.
\end{align}
Intuitively, this is because $\Phi$ measures the increase in the size of the $S^2$ as one climbs out of a black hole's throat towards its asymptotically flat region.  In particular, notice that when the Rindler anabasis solution $\Phi=2(\rho+\kappa)$ is mapped to $\Phi=-2\kappa rt$ via \eqref{slow nearNHEK transfn}, this leads to the range $rt\leq-1$ shown in Fig.~\ref{fig1}.

\section{General Poincar\'e to Rindler transformation}
\label{sec: general PtoR  transformation}

The transformation between the Poincar\'e and Rindler backgrounds used in the previous section [Eq.~\eqref{slow nearNHEK transfn}] maps the Rindler anabasis solution $h_{\theta\theta}=2(\rho+\kappa)$ to the $\mu>0$ perturbation $\Phi=-\sqrt{\mu}\,rt$ in Poincar\'e coordinates \eqref{Phi mu>0 standard} with $\sqrt{\mu}=2\kappa$.  As explained in Section~\ref{subsection: SL(2)}, the standard $\Phi=-\sqrt{\mu}\,rt$ solution is related by \sltwo transformations to any \sltwo-breaking solution $\Phi$ \eqref{Phi soln} with the same $\mu>0$.  As a result, there is a two-parameter generalization of the standard Poincar\'e to Rindler transformation \eqref{slow nearNHEK transfn} that can map the $h_{\theta\theta}=2(\rho+\kappa)$ Rindler anabasis solution to a general Poincar\'e solution $\Phi$ \eqref{Phi soln} with $\sqrt{\mu}=2\kappa$.

On the Penrose diagram of $AdS_2\times S^2$, the two-parameter generalization of Fig.~\ref{fig1} allows for the Rindler patch to have arbitrary vertical location and size with respect to the Poincar\'e one. This is shown in Fig.~\ref{fig2}. The general coordinate transformation is
\begin{align}
\label{general PtoR transfn diffeo}
\begin{aligned}
	t&=\pa{1+\nu^2}\frac{e^{\kappa\tau}\sqrt{\rho(\rho+2\kappa)}\pa{\rho+\kappa+\psi e^{\kappa\tau}\sqrt{\rho(\rho+2\kappa)}}}{\pa{\rho+\kappa+\psi e^{\kappa\tau}\sqrt{\rho(\rho+2\kappa)}}^2-\kappa^2}-\nu\,,\\
	r&=\frac{1}{\pa{1+\nu^2}\kappa}\pa{e^{-\kappa\tau}\sqrt{\rho(\rho+2\kappa)}\pa{1+\psi^2e^{2\kappa\tau}}+2\psi(\rho+\kappa)}\,,
\end{aligned}
\end{align}
accompanied by $A\to A+d\Lambda$,
\begin{align}
\label{general PtoR transfn gauge}
\begin{aligned}
	\Lambda&=-\frac{1}{2}\ln\frac{\rho}{\rho+2\kappa}+\ln\frac{\rho+\psi e^{\kappa\tau}\sqrt{\rho(\rho+2\kappa)}}{\rho+2\kappa+\psi e^{\kappa\tau}\sqrt{\rho(\rho+2\kappa)}}\\
	&=\frac{1}{2}\ln\frac{\br{r(t+\nu)-1}\br{\psi r(t+\nu)-\pa{1+\nu^2}r-\psi}}{\br{r(t+\nu)+1}\br{\psi r(t+\nu)-\pa{1+\nu^2}r+\psi}}\,,
\end{aligned}
\end{align}
with $\psi=\nu-\chi\geq0$.  The derivation of this general transformation is in Appendix~\ref{app: general PtoR  transformation}.

\begin{figure}[!ht]
	\centering
	\resizebox{!}{0.4\textheight}
	{
	\begin{tikzpicture}
		\draw[thick] (0,-1) -- (0, 17);
		\draw[thick] (-8,-1) -- (-8, 17);

		\draw[thick] (0,0) to (-8, 8) to (0,16);

		\draw[thick] (0,4) to (-5, 9) to (0,14);

		\draw (0, 4) node[anchor=west]{$t=-\nu$};
		\draw (0, 14) node[anchor=west]{$t=\frac{1+\nu\chi}{\nu-\chi}$};

		\draw (-5.75, 9.75) node[rotate=45]{$r=0$};
		\draw (-6.4, 7) node[rotate=-45]{$r=0$};
		\draw (-0.3, 2.25) node[rotate=90]{$r=\infty$};

		\draw (-1.75, 11.75) node[rotate=45]{$\rho=0$};	
		\draw (-1.75,6.25) node[rotate=-45]{$\rho=0$};
		\draw (-0.3, 9) node[rotate=90]{$\rho=\infty$};
	\end{tikzpicture}
	}
	\caption{Penrose diagram of $AdS_2\times S^2$ with relative placement of the Poincar\'e patch \eqref{AdS2xS2} and the Rindler patch \eqref{NAdS2xS2} according to the transformation \eqref{general PtoR transfn diffeo}.}
	\label{fig2}
\end{figure}

Using the above general transformation we may ask again: when is it possible to map a Poincar\'e solution $\Phi=ar+brt+cr\left(t^2-1/r^2\right)$ to the Rindler anabasis solution $\Phi=2(\rho+\kappa)$?  We find that the answer is again: when and only when $\mu=b^2-4ac>0$.  For $\mu>0$ we find that the parameter identification is 
\begin{align}
	\kappa=\sqrt{b^2-4ac}/2
	=\sqrt{\mu}/2\,,
\end{align}
and
\begin{align}
	\nu&=\frac{b-\sqrt{b^2-4ac}}{2c}\,,
	&&\chi=\frac{a+c}{\sqrt{b^2-4ac}}\,,
	&&&&\text{for}\quad
	c<0\,,\\
	\nu&=a/b\,,
	&&\chi=a/b\,,
	&&&&\text{for}\quad
	c=0\,,b>0\,,\\
	\nu&=+\infty\,,
	&&\chi=-a/b\,,
	&&&&\textrm{for}\quad
	c=0\,,b<0\,.
\end{align}
Notice that the above does not include any solutions with $c>0$.  The reason is the following.  As noted in \eqref{Phi>0}, only solutions with $\Phi>0$ may be used for anabasis to an asymptotically flat black hole region.  For $c>0$, near the boundary $r\to\infty$, the general Poincar\'e solution is given by $\Phi\approx r \pa{a+bt+ct^2}$, which is positive in a portion of the boundary that has the form $(-\infty,t_1)\cup(t_2,+\infty)$.  As a result, no $\mu>0$ solution $\Phi$ with $c>0$ may be transformed to the Rindler anabasis solution $\Phi=2(\rho+\kappa)$ by a single transformation (\ref{general PtoR transfn diffeo}--\ref{general PtoR transfn gauge}) everywhere near the boundary.  That said, if we allow for topology change when $c>0$, it may be possible to perform anabasis to two separate asymptotically flat regions.

\section{Backreaction calculation in the connected throat}
\label{sec: backreaction calcn}

In the previous sections, we have seen that the \ads backreaction problem for an \sltwo-breaking electrovacuum solution $\Phi$ in four-dimensional Einstein-Maxwell theory is consistent as long as one does not insist on maintaining \ads boundary conditions but rather considers such linear solutions as black hole anabasis solutions that build the asymptotically flat regions of extreme (for $\mu=0$) and near-extreme (for $\mu>0$) Reissner-Nordstr\"om.  In this section, we show that the addition of anabasis perturbations to \ads may also be thought of as a boundary condition for a \emph{connected} $AdS_2$.  The connected \ads is a nearly-\ads with its \sltwo broken appropriately for it to maintain connection to the asymptotically flat region of a Reissner-Nordstr\"om black hole. 

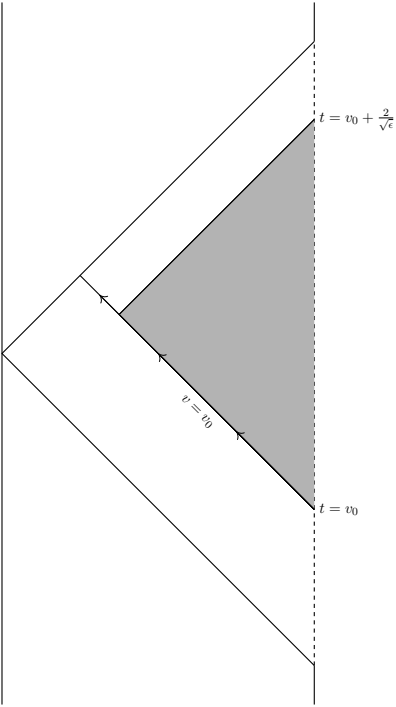
\begin{figure}[!ht]
	\centering
	\resizebox{!}{0.4\textheight}
	{
	\begin{tikzpicture}
		\draw[thick] (0,-1) -- (0, 0);
		\draw[thick] (0,16) -- (0, 17);
		\draw[thick,dashed] (0,0) -- (0, 16);	

		\draw[thick] (-8,-1) -- (-8, 17);

		\draw[thick] (0,0) to (-8, 8) to (0,16);

		\filldraw[fill=black!30] (0,4) to (-5, 9) to (0,14);
		\draw[thick] (0,4) to (-6, 10);
		\draw[thick] (-5, 9) to (0,14);

		\draw[-{>[scale=2.5, length=2, width=3]}] (0,4) to (-5.5, 9.5);

		\draw[-{>[scale=2.5, length=2, width=3]}] (0,4) to (-4, 8);

		\draw[-{>[scale=2.5, length=2, width=3]}] (0,4) to (-2, 6);

		\draw (0, 4) node[anchor=west]{$t=v_0$};
		\draw (0, 14) node[anchor=west]{$t=v_0+\frac{2}{\sqrt{\epsilon }}$};

		\draw (-3,6.5) node[rotate=-45]{$v=v_0$};
	\end{tikzpicture}
	}
	\caption{Penrose diagram of $AdS_2\times S^2$.  The dashed boundary signifies that this \ads is maintaining connection with the exterior of Reissner-Nordstr\"om.  An ingoing pulse of energy $\epsilon$ enters the extreme throat (Poincar\'e patch) at $t=v_0$.  The black hole horizon shifts and the throat becomes near-extreme (shaded Rindler patch).}
	\label{figV}
\end{figure}

Consider, for example, throwing a matter pulse of energy $\epsilon>0$ into the connected \ads throat of an extreme Reissner-Nordstr\"om.  That is, add to the right hand side of the Einstein equation in \eqref{Einstein-Maxwell} the matter energy-momentum tensor
\begin{align}
	\label{matter Tmunu}
	8\pi T_{vv}^{\textrm{matter}}=\epsilon\delta(v-v_0)\,,\quad
	v=t-1/r\,.
\end{align}
For $\epsilon\ll 1$ we may find an $\mathcal{O}(\epsilon)$ metric and gauge field perturbation around the Bertotti-Robinson universe that generalizes the solution (\ref{hthth soln}--\ref{ftr soln}) according to
\begin{align}
	\label{hthth soln epsilon}
	h_{\theta\theta}&=\Phi_0+ar+brt+cr\pa{t^2-1/r^2}-\frac{\epsilon}{2}\frac{r^2(t-v_0)^2-1}{r}\Theta(v-v_0)\,,\\
	\label{htr soln  epsilon}
	h_{tr}&=-\frac{1}{2}rt\br{\Phi_0+2ar+brt+\frac{2}{3}cr\pa{t^2-9/r^2}}\\
	&\quad\,+\frac{\epsilon}{2}\frac{r^3(t-v_0)^3-9r(t-v_0)+6\ln(r(t-v_0))+8}{3r}\Theta(v-v_0)\,,\notag\\
	\label{ftr soln  epsilon}
	f_{tr}&=\p_t a_r
	=h_{\theta\theta}-{\Phi_0/2}\,.
\end{align}
Before the pulse, for $v<v_0$, we impose the causal boundary condition for a connected \ads throat given by the $\mu=0$ Poincar\'e anabasis solution $\Phi=2r$, 
\begin{align}
	\label{bc}
	h_{\theta\theta}=2r\,,\quad
	\text{for}\quad
	v<v_0\,.
\end{align}
That is to say, we set $a=2$ and $b=c=\Phi_0=0$.  Then after the pulse, for $v>v_0$, we get
\begin{align}
	h_{\theta\theta}=2r-\frac{\epsilon}{2}\frac{r^2(t-v_0)^2-1}{r}\,,\quad
	\text{for}\quad
	v>v_0\,.
\end{align}
This solution after the pulse is a $\mu=4\epsilon$ solution which, using the results from Sec.~\ref{sec: general PtoR  transformation}, maps to the Rindler anabasis solution $h_{\theta\theta}=2(\rho+\kappa)$ via (\ref{general PtoR transfn diffeo}--\ref{general PtoR transfn gauge}) with  
\begin{align}
	\kappa=\sqrt{\epsilon}\,,\quad
	\nu=\frac{2}{\sqrt{\epsilon}}-v_0\,,\quad
	\chi=\frac{1}{\sqrt{\epsilon}}-\frac{1+v_0^2}{4}\sqrt{\epsilon}\,.
\end{align}
We thus see that we have a backreaction calculation in the connected \ads throat that is consistent with the expectation from the physics of Reissner-Nordstr\"om:  Throwing a pulse of energy $\epsilon\ll 1$ into the extreme black hole with $Q=M$ ``shifts the horizon'' and the black hole becomes near extreme with $Q=M\sqrt{1-\epsilon}$. This is shown in Fig.~\ref{figV}.

\section{Discussion}
\label{sec: conclusion}

In this paper, we have studied backreaction in the context of $AdS_2\times S^2$ connected to an asymptotically flat region in four-dimensional Einstein-Maxwell theory.  We imposed spherical symmetry but considered both electrovacuum solutions as well as a matter source in the form of a null ingoing pulse.  We have seen that backreaction with boundary condition change, which we call anabasis, is consistent with Reissner-Nordstr\"om physics. 

In AdS/CFT, anabasis is dual to following the inverse renormalization group (RG) flow, from IR to UV, for an appropriate irrelevant deformation of the boundary field theory that does not respect AdS boundary conditions.  This is not something discussed very often in the AdS/CFT literature for at least two reasons.  First, it is a difficult question to study systematically because it is hard to identify appropriate solvable irrelevant deformations of CFTs.  Second, it runs somewhat contrary to the spirit of AdS/CFT which is a complete self-contained theory in itself---a theory in which even when one studies irrelevant deformations, one may wish to restrict oneself to deformations which do not destroy the boundary of AdS.  Historically, of course, AdS/CFT was discovered by a low-energy near-horizon limit from string theory in asymptotically flat spacetime.  A recent body of work that carries out an anabasis by following a flow for a single-trace irrelevant deformation of a CFT$_2$, which goes under the name $T\overline{T}$ and changes $AdS_3$ asymptotics to flat with a linear dilaton, may be found in \cite{Giveon:2017nie,Giveon:2017myj,Asrat:2017tzd,Giribet:2017imm}.\footnote{A different double-trace $T\overline{T}$ deformation of CFT$_2$ has been holographically interpreted as a gravitational theory in an $AdS_3$ that is cut off at a finite interior surface \cite{McGough:2016lol} (see also \cite{Kraus:2018xrn}).  This is not related to anabasis as it may be obtained from mixed boundary conditions that respect the $AdS_3$ boundary \cite{Guica:2019nzm}.}
For $AdS_2$, it is not clear to us how an inverse RG flow would be implemented microscopically due to the apparent lack of irrelevant deformations of quantum mechanical models with conformal symmetry.  For an RG flow along a relevant deformation, useful for flowing away from $AdS_2$ but not for an anabasis from it, see, e.g., \cite{Anninos:2020cwo}.

The gravitational aspects of \ads anabasis studied in this paper do not rely on the existence of a holographic dual and are expected to be readily generalizable to a wide class of theories with (near-)extreme black holes which universally exhibit $AdS_2$-like near-horizon geometries \cite{Kunduri:2007vf}.  This includes rotating black holes such as Kerr which near extremality has a throat geometry, the Near-Horizon-Extreme-Kerr (NHEK) solution \cite{Bardeen:1999px}, with backreaction properties similar to $AdS_2\times S^2$ \cite{Amsel:2009ev,Dias:2009ex}.  In Appendix~\ref{app: NHEK triplet}, we give the \sltwo-breaking triplet of linear perturbations of NHEK that generalizes (\ref{hthth soln}--\ref{ftr soln}).  Beyond near-horizon approximations, \ads makes an appearance in other contexts where approximate spacetime decoupling occurs, such as the interaction region of colliding shock electromagnetic plane waves \cite{Bell:1974vb}, or near certain highly localized matter distributions \cite{Meinel:2011ur}.  The ideas in this paper may also be relevant in such contexts.

A model of two-dimensional dilaton gravity in \ads that is solvable with backreaction, as well as with the addition of matter, is the Jackiw-Teitelboim (JT) theory \cite{Teitelboim:1983ux,Jackiw:1984je}.  This model captures many of the universal aspects in the spherically symmetric sector of higher dimensional gravity near extreme black hole horizons, and it has been studied extensively from the holographic perspective beginning with \cite{Almheiri:2014cka,Jensen:2016pah,Maldacena:2016upp,Engelsoy:2016xyb}.  In JT theory, the geometry is fixed to being locally \ads but the \sltwo is broken by a dilaton $\Phi_{JT}$.  Comparing with our gravitational perturbations, we may identify $\Phi=\Phi_{JT}$, noting that \eqref{Phi soln} solves the JT equation of motion $\nabla_\mu\nabla_\nu\Phi_{JT}-g_{\mu\nu}\nabla^2\Phi_{JT}+g_{\mu\nu}\Phi_{JT}=0$ for \ads in Poincar\'e coordinates.  This is because in the ansatz \eqref{pertn ansatz} we have $\Phi$ measuring the variation in the size of the $S^2$ and, in this ansatz, dimensional reduction of higher dimensional gravity down to two dimensions is known to lead to JT theory with the dilaton $\Phi_{JT}$ measuring precisely this variation (see e.g \cite{Almheiri:2016fws,Nayak:2018qej,Moitra:2018jqs,Sachdev:2019bjn}).  Continuing the comparison, the \sltwo-invariant $\mu$ defined in \eqref{mu defn} may be identified with the ADM mass of the 2D black holes in JT theory \cite{Almheiri:2014cka,Mann:1992yv}.  It follows that the mass of the 2D \ads black hole in JT is the deviation from extremality of the 4D Reissner-Nordstr\"om in Einstein-Maxwell.  A comment is in order here.  It is often said in the literature that JT is a nearly-\ads theory with the ``nearly'' part, which is due to the dilaton's breaking of the \sltwo symmetry of $AdS_2$, associated with a departure from extremality.  As we have seen in this paper, however, this is not necessarily so because for $\mu=0$ the \sltwo may be broken only in order to build the exterior of an exactly extreme Reissner-Nordstr\"om.

The connected $AdS_2\times S^2$, which we defined in Section~\ref{sec: backreaction calcn} in order to perform a consistent Reissner-Nordstr\"om backreaction calculation, is nearly-\ads in the sense that its \sltwo has been broken by the addition of anabasis perturbations that make this \ads an approximate one.  We also saw that this \sltwo breaking may be thought of as a choice of boundary condition for the backreaction calculation.  A comprehensive study of various boundary conditions for the JT theory has been carried out in \cite{Goel:2020yxl}.  However, it appears that none of the boundary conditions contained therein would yield an anabasis as they at best correspond to mixed boundary conditions that do not destroy the \ads boundary (of the double-trace type in AdS/CFT terms).  On the other hand, the boundary term used in \cite{Brown:2018bms} for what is called therein ``permeable boundary conditions'' appears to be a better candidate for defining a connected \ads in JT theory.  Indeed, matching fields across the $AdS_2\times S^2$ boundary in Reissner-Nordstr\"om, as in the calculations of \cite{Porfyriadis:2018yag,Porfyriadis:2018jlw}, necessitates boundary conditions that are ``leaky'' from the \ads point of view. 

Broadening the pulse used in Section~\ref{sec: backreaction calcn} and replacing it with a finite-width wavepacket, one may arrange to have such a wavepacket enter the $AdS_2\times S^2$ region by sending in low energy waves from past null infinity in Reissner-Nordstr\"om.  The calculation may be set up using matched asymptotic expansions and features leaky boundary conditions \cite{mae:paper}.

Generically, the backreaction of an extreme Reissner-Nordstr\"om black hole results, as in Section~\ref{sec: backreaction calcn}, in a near-extreme one.  However, there is a notable exception.  In \cite{Murata:2013daa} it was found that there exist fine-tuned initial data for a massless scalar perturbing extreme Reissner-Nordstr\"om, for which an instability of the scalar field at the event horizon --the Aretakis instability-- persists for arbitrarily long evolution and leads to a spacetime that may be thought of as a dynamical extreme black hole.  It was observed that, at late times, this dynamical extreme black hole has the same exterior as extreme Reissner-Nordstr\"om but differs from it at the horizon.  In \cite{Hadar:2018izi}, the Aretakis instability of the perturbing massless scalar on extreme Reissner-Nordstr\"om was analyzed using the symmetries of its $AdS_2\times S^2$ throat.  It would be interesting to study the dynamical extreme black hole of \cite{Murata:2013daa} using a connected $AdS_2\times S^2$ as defined in this paper.

\acknowledgements

We thank Andrew Strominger for helpful discussions.  This work was supported by the Black Hole Initiative at Harvard University, which is funded by grants from the John Templeton Foundation and the Gordon and Betty Moore Foundation.  SH and AL gratefully acknowledge support from the Jacob Goldfield Foundation.  Funding for shared facilities used in this research was provided by NSF Grant No. 1707938.

\appendix

\section{\boldmath Derivation of the transformation in Sec.~\ref{sec: general PtoR  transformation}}
\label{app: general PtoR  transformation}

In this appendix, we give some details about the derivation of the general Poincar\'e to Rindler transformation (\ref{general PtoR transfn diffeo}--\ref{general PtoR transfn gauge}) in Sec.~\ref{sec: general PtoR  transformation}.  In particular, we show how this general transformation may be derived by composing a global $AdS_2$ time translation with a Poincar\'e time translation of the basic transformation \eqref{slow nearNHEK transfn}.

\begin{figure}[!ht]
	\resizebox{!}{0.4\textheight}
	{
 	\begin{tikzpicture}
		\draw[thick] (0,-1) -- (0, 21);
		\draw[thick] (-8,-1) -- (-8, 21);

		\draw[thick] (0,0) to (-8, 8) to (0,16);

		\draw[thick] (0,4) to (-8, 12) to (0,20);

		\draw (0, 4) node[anchor=west]{$t=-\nu$};
		\draw (0, 16) node[anchor=west]{$\tilde{t}=\nu$};

		\draw (-3.75, 11.75) node[rotate=45]{$r=0$};
		\draw (-3.75, 4.25) node[rotate=-45]{$r=0$};
		\draw (-0.3, 2.25) node[rotate=90]{$r=\infty$};

		\draw (-2.75, 16.75) node[rotate=45]{$\tilde{r}=0$};
		\draw (-2.75,7.25) node[rotate=-45]{$\tilde{r}=0$};
		\draw (-0.3, 17.75) node[rotate=90]{$\tilde{r}=\infty$};
	\end{tikzpicture}
	}
	\hspace{0.1\textheight}
	\resizebox{!}{0.4\textheight}
	{
	\begin{tikzpicture}
		\draw[thick] (0,-1) -- (0, 21);
		\draw[thick] (-8,-1) -- (-8, 21);

		\draw[thick] (0,4) to (-8, 12) to (0,20);

		\draw[thick] (0,4) to (-5.25, 9.25) to (0,14.5);
	
		\draw (0, 14.5) node[anchor=west]{$\tilde{t}=\chi$};

		\draw (-5, 14.5) node[rotate=45]{$\tilde{r}=0$};
		\draw (-6.25, 10.75) node[rotate=-45]{$\tilde{r}=0$};
		\draw (-0.3, 17) node[rotate=90]{$\tilde{r}=\infty$};

		\draw (-2.5, 11.5) node[rotate=45]{$\rho=0$};	
		\draw (-2.5, 7) node[rotate=-45]{$\rho=0$};
		\draw (-0.3, 9.25) node[rotate=90]{$\rho=\infty$};
	\end{tikzpicture}
	}
	\caption{Penrose diagrams of $AdS_2\times S^2$.  Left: The two Poincar\'e patches are related by a global time translation according to the transformation \eqref{global AdS2 time trasln diffeo}.  Right: The Poincar\'e and Rindler patches are situated according to the transformation \eqref{Poincare AdS2 time trasln diffeo}.}
	\label{figApp}
\end{figure}
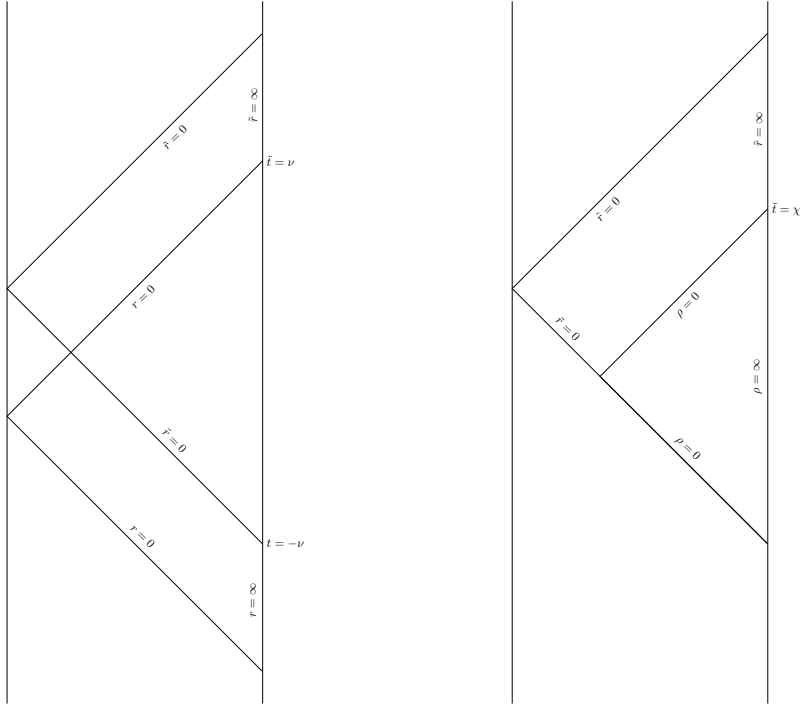

The global $AdS_2$ coordinates, with $ds_2^2=(-d\eta^2+d\sigma^2)/\sin^2{\sigma}$, are related to the Poincar\'e coordinates via $t\pm\frac{1}{r}=\tan\pa{\frac{\eta\pm\sigma}{2}}$.  The translation $\eta\to\eta+\eta_0$ then moves the Poincar\'e patch vertically on the Penrose diagram of $AdS_2$ as shown in the left panel of Fig.~\ref{figApp}.  In the overlapping region, with $ds_2^2=-r^2dt^2+dr^2/r^2=-\tilde{r}^2d\tilde{t}^2+d\tilde{r}^2/\tilde{r}^2$, the relation between the two different sets of Poincar\'e coordinates reads
\begin{align}
\label{global AdS2 time trasln diffeo}
\begin{aligned}
	t&=-\pa{1+\nu^2}\frac{\tilde{r}^2\pa{\tilde{t}-\nu}}{\tilde{r}^2\pa{\tilde{t}-\nu}^2-1}-\nu\,,\\
	r&=\frac{1}{1+\nu^2}\frac{\tilde{r}^2\pa{\tilde{t}-\nu}^2-1}{\tilde{r}}\,,
\end{aligned}
\end{align}
with $\nu=\cot(\eta_0/2)$.  For the Bertotti-Robinson solution, the above coordinate transformation must be supplemented by the gauge field transformation $A\to A+d\Lambda\,,\Lambda=\ln\frac{r(t+\nu)-1}{r(t+\nu)+1}$.

The resizing of the Rindler patch situated as in Fig.~\ref{fig1} is achieved by a translation of the Poincar\'e time coordinate in the transformation \eqref{slow nearNHEK transfn}.  Specifically, the transformation
\begin{align}
\label{Poincare AdS2 time trasln diffeo}
\begin{aligned}
	\tilde{t}&=-e^{-\kappa\tau}\frac{\rho+\kappa}{\sqrt{\rho(\rho+2\kappa)}}+\chi\,,\\
	\tilde{r}&=\frac{1}{\kappa}e^{\kappa\tau}\sqrt{\rho(\rho+2\kappa)}\,,
\end{aligned}
\end{align}
together with $A\to A+d\Lambda\,,\Lambda=\frac{1}{2}\ln\frac{\tilde{r}\pa{\tilde{t}-\chi}-1}{\tilde{r}\pa{\tilde{t}-\chi}+1}$ implements the mapping shown in the right panel of Fig.~\ref{figApp}.  

The transformation (\ref{general PtoR transfn diffeo}--\ref{general PtoR transfn gauge}) is a composition of \eqref{global AdS2 time trasln diffeo} with \eqref{Poincare AdS2 time trasln diffeo}.  Note that one must have $\nu\geq\chi$ in order for the Rindler patch to be contained entirely inside the Poincar\'e one covered by the $(t,r)$ coordinates.

\section{\texorpdfstring{\boldmath\sltwo}{SL(2)}-breaking NHEK perturbations}
\label{app: NHEK triplet}

The Near-Horizon of Extreme-Kerr (NHEK) was obtained in \cite{Bardeen:1999px} by a scaling limit $\lambda\to0$ applied to extreme Kerr, analogous to the one that produces Bertotti-Robinson from extreme Reissner-Nordstr\"om.  The $\mathcal{O}(1)$ NHEK metric is given by
\begin{gather}
	ds^2=2M^2\Gamma\br{-r^2dt^2+\frac{dr^2}{r^2}+d\theta^2+\Lambda^2\pa{d\phi+rdt}^2},\\
	\Gamma(\theta)=\frac{1+\cos^2{\theta}}{2},\quad
	\Lambda(\theta)=\frac{2\sin{\theta}}{1+\cos^2{\theta}}.
\end{gather}
NHEK has an isometry group given by $\mathsf{SL}(2)\times\mathsf{U}(1)$, where the $\mathsf{U}(1)$ is due to axial symmetry and the $\mathsf{SL}(2)$ is again associated with the \ads part in the above. 

An \sltwo-breaking triplet of axially symmetric linear perturbations of NHEK that generalizes (\ref{hthth soln}--\ref{ftr soln}) is given by
\begin{gather}
	h_{tt}=4\Gamma\p_t^2\Phi+r^2\pa{2\Gamma-1}\pa{1+\Lambda^2}\Phi+4rA_t\Gamma\Lambda^2,\quad
	h_{tr}=2rA_r\Gamma\Lambda^2,\\
	h_{t\phi}=\Phi r\pa{2\Gamma-1}\Lambda^2+2A_t\Gamma\Lambda^2,\quad
	h_{rr}=\frac{\Phi}{r^2},\quad
	h_{r\phi}=2A_r\Gamma\Lambda^2,\\
	h_{\theta\theta}=\Phi,\quad
	h_{\phi\phi}=\Phi\pa{2\Gamma-1}\Lambda^2,
\end{gather}
with
\begin{gather}
	A_t=-r^2\frac{\Gamma^2\Lambda^2}{8}\p_r\Phi,\quad
	A_r=\frac{1}{r^2}\pa{1-\frac{\Gamma^2\Lambda^2}{8}}\p_t\Phi,
\end{gather}
and $\Phi$ as in \eqref{Phi soln}.  We derived the above as follows.  First, we obtained the anabasis solution, corresponding to $\Phi=2r$, from the $\mathcal{O}(\lambda)$ term in the expansion of \cite{Bardeen:1999px}.  Then, we applied \sltwo transformations to generate the triplet while adjusting the gauge for clarity.  We note that for NHEK perturbations, even within axial symmetry, $h_{\theta\theta}$ is not gauge-invariant.

We expect the anabasis of these three perturbations towards Kerr to proceed in a similar fashion to the analysis carried out above for Reissner-Nordstr\"om.  However, it is worth emphasizing that there is no analog of Birkhoff's theorem for axisymmetric spacetimes and that there exist axisymmetric propagating gravitational wave perturbations of Kerr and NHEK.  These are typically studied in the Newman–Penrose formalism as in \cite{Amsel:2009ev,Dias:2009ex}.  In \cite{Castro:2019crn} (see also \cite{Castro:2018ffi}) an attempt was made to find NHEK perturbations using a metric ansatz judiciously picked to accommodate the anabasis perturbation to (near-)extreme Kerr.  Unfortunately, the solutions found in \cite{Castro:2019crn} that go beyond the above triplet are singular at the poles $\theta=0,\pi$.\footnote{In their notation, $\Psi$ diverges at the poles unless $\chi=\Phi$, in which case their \sltwo-breaking solution reduces to the above triplet.}

Finally, note that in another gauge, our solution triplet takes the simple form
\begin{align}
	h_{\mu\nu}&=2M^2\Gamma
	\begin{pmatrix}
		r^2\Phi\Gamma & h_{tr} & 0 & 0 \\
		& \frac{\Phi\pa{2-\Gamma}}{r^2} & 0 & 0 \\
		& & \Phi\Gamma^2\Lambda^2 & 0 \\
		& &	& 0
	\end{pmatrix}
	\,,\\
	h_{tr}&=\frac{1}{3}rt\br{2ar+brt+\frac{2}{3}cr\pa{t^2+9/r^2}}.
\end{align}

\bibliography{anabasis.bib}

\providecommand{\href}[2]{#2}\begingroup\raggedright\begin{thebibliography}{10}

\bibitem{Hawking:1973uf}
S.~Hawking and G.~Ellis, \href{http://dx.doi.org/10.1017/CBO9780511524646}{{\em
  {The Large Scale Structure of Space-Time}}}.
\newblock Cambridge Monographs on Mathematical Physics. Cambridge University
  Press, 2, 2011.

\bibitem{Maldacena:1998uz}
J.~M. Maldacena, J.~Michelson, and A.~Strominger, ``{Anti-de Sitter
  fragmentation},'' \href{http://dx.doi.org/10.1088/1126-6708/1999/02/011}{{\em
  JHEP} {\bfseries 02} (1999) 011},
  \href{http://arxiv.org/abs/hep-th/9812073}{{\ttfamily arXiv:hep-th/9812073}}.

\bibitem{Murata:2013daa}
K.~Murata, H.~S. Reall, and N.~Tanahashi, ``{What happens at the horizon(s) of
  an extreme black hole?},''
  \href{http://dx.doi.org/10.1088/0264-9381/30/23/235007}{{\em Class. Quant.
  Grav.} {\bfseries 30} (2013) 235007},
  \href{http://arxiv.org/abs/1307.6800}{{\ttfamily arXiv:1307.6800 [gr-qc]}}.

\bibitem{Spradlin:1999bn}
M.~Spradlin and A.~Strominger, ``{Vacuum states for AdS$_{2}$ black holes},''
  \href{http://dx.doi.org/10.1088/1126-6708/1999/11/021}{{\em JHEP} {\bfseries
  11} (1999) 021}, \href{http://arxiv.org/abs/hep-th/9904143}{{\ttfamily
  arXiv:hep-th/9904143}}.

\bibitem{Giveon:2017nie}
A.~Giveon, N.~Itzhaki, and D.~Kutasov, ``{$ \mathrm{T}\overline{\mathrm{T}} $
  and LST},'' \href{http://dx.doi.org/10.1007/JHEP07(2017)122}{{\em JHEP}
  {\bfseries 07} (2017) 122}, \href{http://arxiv.org/abs/1701.05576}{{\ttfamily
  arXiv:1701.05576 [hep-th]}}.

\bibitem{Giveon:2017myj}
A.~Giveon, N.~Itzhaki, and D.~Kutasov, ``{A solvable irrelevant deformation of
  AdS$_{3}$/CFT$_{2}$},'' \href{http://dx.doi.org/10.1007/JHEP12(2017)155}{{\em
  JHEP} {\bfseries 12} (2017) 155},
  \href{http://arxiv.org/abs/1707.05800}{{\ttfamily arXiv:1707.05800
  [hep-th]}}.

\bibitem{Asrat:2017tzd}
M.~Asrat, A.~Giveon, N.~Itzhaki, and D.~Kutasov, ``{Holography Beyond AdS},''
  \href{http://dx.doi.org/10.1016/j.nuclphysb.2018.05.005}{{\em Nucl. Phys. B}
  {\bfseries 932} (2018) 241--253},
  \href{http://arxiv.org/abs/1711.02690}{{\ttfamily arXiv:1711.02690
  [hep-th]}}.

\bibitem{Giribet:2017imm}
G.~Giribet, ``{$T\bar{T}$-deformations, AdS/CFT and correlation functions},''
  \href{http://dx.doi.org/10.1007/JHEP02(2018)114}{{\em JHEP} {\bfseries 02}
  (2018) 114}, \href{http://arxiv.org/abs/1711.02716}{{\ttfamily
  arXiv:1711.02716 [hep-th]}}.

\bibitem{McGough:2016lol}
L.~McGough, M.~Mezei, and H.~Verlinde, ``{Moving the CFT into the bulk with $
  T\overline{T} $},'' \href{http://dx.doi.org/10.1007/JHEP04(2018)010}{{\em
  JHEP} {\bfseries 04} (2018) 010},
  \href{http://arxiv.org/abs/1611.03470}{{\ttfamily arXiv:1611.03470
  [hep-th]}}.

\bibitem{Kraus:2018xrn}
P.~Kraus, J.~Liu, and D.~Marolf, ``{Cutoff AdS$_{3}$ versus the $ T\overline{T}
  $ deformation},'' \href{http://dx.doi.org/10.1007/JHEP07(2018)027}{{\em JHEP}
  {\bfseries 07} (2018) 027}, \href{http://arxiv.org/abs/1801.02714}{{\ttfamily
  arXiv:1801.02714 [hep-th]}}.

\bibitem{Guica:2019nzm}
M.~Guica and R.~Monten, ``{$T\bar T$ and the mirage of a bulk cutoff},''
  \href{http://arxiv.org/abs/1906.11251}{{\ttfamily arXiv:1906.11251
  [hep-th]}}.

\bibitem{Anninos:2020cwo}
D.~Anninos and D.~A. Galante, ``{Constructing AdS$_{2}$ flow geometries},''
  \href{http://dx.doi.org/10.1007/JHEP02(2021)045}{{\em JHEP} {\bfseries 02}
  (2021) 045}, \href{http://arxiv.org/abs/2011.01944}{{\ttfamily
  arXiv:2011.01944 [hep-th]}}.

\bibitem{Kunduri:2007vf}
H.~K. Kunduri, J.~Lucietti, and H.~S. Reall, ``{Near-horizon symmetries of
  extremal black holes},''
  \href{http://dx.doi.org/10.1088/0264-9381/24/16/012}{{\em Class. Quant.
  Grav.} {\bfseries 24} (2007) 4169--4190},
  \href{http://arxiv.org/abs/0705.4214}{{\ttfamily arXiv:0705.4214 [hep-th]}}.

\bibitem{Bardeen:1999px}
J.~M. Bardeen and G.~T. Horowitz, ``{The Extreme Kerr throat geometry: A Vacuum
  analog of AdS$_{2}\times S^2$},''
  \href{http://dx.doi.org/10.1103/PhysRevD.60.104030}{{\em Phys. Rev. D}
  {\bfseries 60} (1999) 104030},
  \href{http://arxiv.org/abs/hep-th/9905099}{{\ttfamily arXiv:hep-th/9905099}}.

\bibitem{Amsel:2009ev}
A.~J. Amsel, G.~T. Horowitz, D.~Marolf, and M.~M. Roberts, ``{No Dynamics in
  the Extremal Kerr Throat},''
  \href{http://dx.doi.org/10.1088/1126-6708/2009/09/044}{{\em JHEP} {\bfseries
  09} (2009) 044}, \href{http://arxiv.org/abs/0906.2376}{{\ttfamily
  arXiv:0906.2376 [hep-th]}}.

\bibitem{Dias:2009ex}
O.~J. Dias, H.~S. Reall, and J.~E. Santos, ``{Kerr-CFT and gravitational
  perturbations},'' \href{http://dx.doi.org/10.1088/1126-6708/2009/08/101}{{\em
  JHEP} {\bfseries 08} (2009) 101},
  \href{http://arxiv.org/abs/0906.2380}{{\ttfamily arXiv:0906.2380 [hep-th]}}.

\bibitem{Bell:1974vb}
P.~Bell and P.~Szekeres, ``{Interacting electromagnetic shock waves in general
  relativity},'' \href{http://dx.doi.org/10.1007/BF00770217}{{\em Gen. Rel.
  Grav.} {\bfseries 5} (1974) 275--286}.

\bibitem{Meinel:2011ur}
R.~Meinel and M.~Hutten, ``{On the black hole limit of electrically
  counterpoised dust configurations},''
  \href{http://dx.doi.org/10.1088/0264-9381/28/22/225010}{{\em Class. Quant.
  Grav.} {\bfseries 28} (2011) 225010},
  \href{http://arxiv.org/abs/1105.3807}{{\ttfamily arXiv:1105.3807 [gr-qc]}}.

\bibitem{Teitelboim:1983ux}
C.~Teitelboim, ``{Gravitation and Hamiltonian Structure in Two Space-Time
  Dimensions},'' \href{http://dx.doi.org/10.1016/0370-2693(83)90012-6}{{\em
  Phys. Lett. B} {\bfseries 126} (1983) 41--45}.

\bibitem{Jackiw:1984je}
R.~Jackiw, ``{Lower Dimensional Gravity},''
  \href{http://dx.doi.org/10.1016/0550-3213(85)90448-1}{{\em Nucl. Phys. B}
  {\bfseries 252} (1985) 343--356}.

\bibitem{Almheiri:2014cka}
A.~Almheiri and J.~Polchinski, ``{Models of AdS$_{2}$ backreaction and
  holography},'' \href{http://dx.doi.org/10.1007/JHEP11(2015)014}{{\em JHEP}
  {\bfseries 11} (2015) 014}, \href{http://arxiv.org/abs/1402.6334}{{\ttfamily
  arXiv:1402.6334 [hep-th]}}.

\bibitem{Jensen:2016pah}
K.~Jensen, ``{Chaos in AdS$_2$ Holography},''
  \href{http://dx.doi.org/10.1103/PhysRevLett.117.111601}{{\em Phys. Rev.
  Lett.} {\bfseries 117} no.~11, (2016) 111601},
  \href{http://arxiv.org/abs/1605.06098}{{\ttfamily arXiv:1605.06098
  [hep-th]}}.

\bibitem{Maldacena:2016upp}
J.~Maldacena, D.~Stanford, and Z.~Yang, ``{Conformal symmetry and its breaking
  in two dimensional Nearly Anti-de-Sitter space},''
  \href{http://dx.doi.org/10.1093/ptep/ptw124}{{\em PTEP} {\bfseries 2016}
  no.~12, (2016) 12C104}, \href{http://arxiv.org/abs/1606.01857}{{\ttfamily
  arXiv:1606.01857 [hep-th]}}.

\bibitem{Engelsoy:2016xyb}
J.~Engels\"oy, T.~G. Mertens, and H.~Verlinde, ``{An investigation of AdS$_{2}$
  backreaction and holography},''
  \href{http://dx.doi.org/10.1007/JHEP07(2016)139}{{\em JHEP} {\bfseries 07}
  (2016) 139}, \href{http://arxiv.org/abs/1606.03438}{{\ttfamily
  arXiv:1606.03438 [hep-th]}}.

\bibitem{Almheiri:2016fws}
A.~Almheiri and B.~Kang, ``{Conformal Symmetry Breaking and Thermodynamics of
  Near-Extremal Black Holes},''
  \href{http://dx.doi.org/10.1007/JHEP10(2016)052}{{\em JHEP} {\bfseries 10}
  (2016) 052}, \href{http://arxiv.org/abs/1606.04108}{{\ttfamily
  arXiv:1606.04108 [hep-th]}}.

\bibitem{Nayak:2018qej}
P.~Nayak, A.~Shukla, R.~M. Soni, S.~P. Trivedi, and V.~Vishal, ``{On the
  Dynamics of Near-Extremal Black Holes},''
  \href{http://dx.doi.org/10.1007/JHEP09(2018)048}{{\em JHEP} {\bfseries 09}
  (2018) 048}, \href{http://arxiv.org/abs/1802.09547}{{\ttfamily
  arXiv:1802.09547 [hep-th]}}.

\bibitem{Moitra:2018jqs}
U.~Moitra, S.~P. Trivedi, and V.~Vishal, ``{Extremal and near-extremal black
  holes and near-CFT$_{1}$},''
  \href{http://dx.doi.org/10.1007/JHEP07(2019)055}{{\em JHEP} {\bfseries 07}
  (2019) 055}, \href{http://arxiv.org/abs/1808.08239}{{\ttfamily
  arXiv:1808.08239 [hep-th]}}.

\bibitem{Sachdev:2019bjn}
S.~Sachdev, ``{Universal low temperature theory of charged black holes with
  AdS$_2$ horizons},'' \href{http://dx.doi.org/10.1063/1.5092726}{{\em J. Math.
  Phys.} {\bfseries 60} no.~5, (2019) 052303},
  \href{http://arxiv.org/abs/1902.04078}{{\ttfamily arXiv:1902.04078
  [hep-th]}}.

\bibitem{Mann:1992yv}
R.~B. Mann, ``{Conservation laws and 2-D black holes in dilaton gravity},''
  \href{http://dx.doi.org/10.1103/PhysRevD.47.4438}{{\em Phys. Rev. D}
  {\bfseries 47} (1993) 4438--4442},
  \href{http://arxiv.org/abs/hep-th/9206044}{{\ttfamily arXiv:hep-th/9206044}}.

\bibitem{Goel:2020yxl}
A.~Goel, L.~V. Iliesiu, J.~Kruthoff, and Z.~Yang, ``{Classifying boundary
  conditions in JT gravity: from energy-branes to $\alpha$-branes},''
  \href{http://arxiv.org/abs/2010.12592}{{\ttfamily arXiv:2010.12592
  [hep-th]}}.

\bibitem{Brown:2018bms}
A.~R. Brown, H.~Gharibyan, H.~W. Lin, {\em et~al.}, ``{Complexity of
  Jackiw-Teitelboim gravity},''
  \href{http://dx.doi.org/10.1103/PhysRevD.99.046016}{{\em Phys. Rev. D}
  {\bfseries 99} no.~4, (2019) 046016},
  \href{http://arxiv.org/abs/1810.08741}{{\ttfamily arXiv:1810.08741
  [hep-th]}}.

\bibitem{Porfyriadis:2018yag}
A.~P. Porfyriadis, ``{Scattering of gravitational and electromagnetic waves off
  $AdS_2\times S^2$ in extreme Reissner-Nordstrom},''
  \href{http://dx.doi.org/10.1007/JHEP07(2018)064}{{\em JHEP} {\bfseries 07}
  (2018) 064}, \href{http://arxiv.org/abs/1805.12409}{{\ttfamily
  arXiv:1805.12409 [hep-th]}}.

\bibitem{Porfyriadis:2018jlw}
A.~P. Porfyriadis, ``{Near-$AdS_2$ perturbations and the connection with
  near-extreme Reissner\textendash{}Nordstrom},''
  \href{http://dx.doi.org/10.1140/epjc/s10052-019-7347-6}{{\em Eur. Phys. J. C}
  {\bfseries 79} no.~10, (2019) 841},
  \href{http://arxiv.org/abs/1806.07097}{{\ttfamily arXiv:1806.07097
  [hep-th]}}.

\bibitem{mae:paper}
S.~Hadar, A.~Lupsasca, and A.~P. Porfyriadis, ``{$AdS_2$ reaction in the
  extreme Reissner-Nordstr\"om throat},''. \emph{In progress}.

\bibitem{Hadar:2018izi}
S.~Hadar, ``{Near-extremal black holes at late times, backreacted},''
  \href{http://dx.doi.org/10.1007/JHEP01(2019)214}{{\em JHEP} {\bfseries 01}
  (2019) 214}, \href{http://arxiv.org/abs/1811.01022}{{\ttfamily
  arXiv:1811.01022 [hep-th]}}.

\bibitem{Castro:2019crn}
A.~Castro and V.~Godet, ``{Breaking away from the near horizon of extreme
  Kerr},'' \href{http://dx.doi.org/10.21468/SciPostPhys.8.6.089}{{\em SciPost
  Phys.} {\bfseries 8} no.~6, (2020) 089},
  \href{http://arxiv.org/abs/1906.09083}{{\ttfamily arXiv:1906.09083
  [hep-th]}}.

\bibitem{Castro:2018ffi}
A.~Castro, F.~Larsen, and I.~Papadimitriou, ``{5D rotating black holes and the
  nAdS$_{2}$/nCFT$_{1}$ correspondence},''
  \href{http://dx.doi.org/10.1007/JHEP10(2018)042}{{\em JHEP} {\bfseries 10}
  (2018) 042}, \href{http://arxiv.org/abs/1807.06988}{{\ttfamily
  arXiv:1807.06988 [hep-th]}}.

\end{thebibliography}\endgroup
\bibliographystyle{utphys}

\end{document}